\documentclass[sigconf]{acmart}
\usepackage{multirow}
\usepackage{subcaption}
\usepackage{algpseudocode}
\usepackage[ruled,linesnumbered]{algorithm2e}
\usepackage{enumitem}
\usepackage{makecell}

\begin{document}
\title{Model Stealing Attack against Recommender System}
\author{Zhihao Zhu, Rui Fan, Chenwang Wu, Yi Yang, Defu Lian, Enhong Chen}

\begin{abstract}
Recent studies have demonstrated the vulnerability of recommender systems to data privacy attacks.
However, research on the threat to model privacy in recommender systems, such as model stealing attacks, is still in its infancy. Some adversarial attacks have achieved model stealing attacks against recommender systems, to some extent, by collecting abundant training data of the target model (target data) or making a mass of queries. 
In this paper, we constrain the volume of available target data and queries and utilize auxiliary data, which shares the item set with the target data, to promote model stealing attacks.
Although the target model treats target and auxiliary data differently, their similar behavior patterns allow them to be fused using an attention mechanism to assist attacks.
Besides, we design stealing functions to effectively extract the recommendation list obtained by querying the target model. 
Experimental results show that the proposed methods are applicable to most recommender systems and various scenarios and exhibit excellent attack performance on multiple datasets.
\end{abstract}

\maketitle

\section{Introduction}
In the contemporary age of information explosion, recommender systems have gained immense popularity in domains like marketing~\cite{ansari2000internet} and e-commerce~\cite{schafer2001commerce, wei2007survey} owing to their exceptional ability to provide personalized recommendations.
By analyzing user-item interactions and external knowledge, recommender systems model user preferences and item attributes, and then recommend items that users are more likely to be interested in.

In recent years, several academic investigations~\cite{lam2006you, zhang2021membership, zhang2022comprehensive} have highlighted the serious privacy risks that recommender systems encounter.
Attackers can exploit their query permissions and other auxiliary knowledge to peek into the training data of recommender systems~\cite{zhang2021membership, zhu2023membership} or privacy attributes of their users.~\cite{zhang2021graph}.
These methods pose serious threats to the security of recommender systems from the perspective of data privacy. However, there is still a lack of relevant research on the model privacy leakage threats faced by recommender systems, such as model stealing attacks~\cite{miura2021megex, takemura2020model, tramer2016stealing}. Model stealing attacks seek to obtain a good copy of the target model, namely the clone model. These attacks can compromise the owner’s intellectual property and increase the risk of adversarial attacks~\cite{madry2017towards, finlayson2019adversarial} and membership inference attacks~\cite{zhang2021membership, wang2022debiasing}, undermining the credibility and privacy of the models. 

\begin{table}[]
\centering
\caption{Our work vs. previous works}
\begin{tabular}{l|c|c}
\hline
Algorithm      & Target data & Query \\
\hline
QSD~\cite{yue2021black}      &  None             &  Abundant \\
Adversarial Attacks~\cite{li2016data, wu2021triple}   &  Abundant   &  None               \\
Our work          &  Few   &  Few                 \\
\hline
\end{tabular}
\label{knowledge}
\end{table}

There is currently no work that formally proposes model stealing attacks against recommender systems. Some articles~\cite{li2016data, wu2021triple} have established surrogate models to achieve adversarial attacks against recommender systems, which have, to some extent, achieved model stealing attacks. However, the aforementioned works frequently assume that the attacker can obtain a large amount of training data for the target model. This assumption is unreasonable in recommender systems because confidentiality and credibility are so important in recommender systems that their owners would pay extra attention to preventing training data exposure. In addition, the attackers did not instruct the surrogate model to replicate the target model, which is the main purpose of model stealing attacks; rather, they employed the surrogate model solely as an intermediate to carry out adversarial attacks. 
QSD(\textbf{Q}uery with \textbf{S}ynthesized \textbf{D}ata)~\cite{yue2021black}, another adversarial attack, makes use of synthetic data rather than target data to construct surrogate models. They synthesized user interactions by utilizing the autoregressive nature of sequential recommender systems, produced a huge number of queries, and then trained a surrogate model by extracting the outputs acquired from the queries. However, the data generation method that they have proposed is only appropriate for sequential recommender systems, and the assumption that a large number of queries will be made is also unreasonable.

To properly define a model stealing attack against recommender systems, we assume that the attacker can only access a small amount of target data and conduct a small number of queries. This assumption is based on the challenges presented by previous research. 
Table~\ref{knowledge} presents a comparison of our study to other works.
Furthermore, in reference to previous privacy attacks~\cite{zhang2021membership, wang2022debiasing} on recommender systems, we hypothesize that the attacker can collect a portion of the auxiliary data, which that comes from the same distribution as the target data but does not take part in the training process of the target model. 
We employ corresponding strategies to exploit three types of knowledge: target data, auxiliary data, and query permission.

First, when an attacker can only access partial target data, we extract the $(user, item)$ interaction pairs from the available target data and encourage the clone model to predict higher ratings for these interaction pairs. We name this attack \textbf{PTD} (\textbf{P}artial \textbf{T}arget \textbf{D}ata). 
It is effective because most recommender systems have the same training objective as ours: to provide higher predicted ratings for interaction pairs in the training data. 
Some works~\cite{li2016data, nguyen2023poisoning} took this way to build the surrogate model for further adversarial attacks against recommender systems.
They assume that attackers can obtain a sufficient amount of target data. In this scenario, PTD is enough to learn a threatening clone model. However, in the real world, most recommender systems do not disclose their complete training data, so we can only improve the performance of model stealing attacks by obtaining other knowledge, including auxiliary data and the recommendation lists of the target model.

Second, when the auxiliary data is available, we utilize an attention mechanism to fuse auxiliary data and available target data. Specifically, we first train an auxiliary model on the auxiliary data and obtain corresponding auxiliary item embedding vectors. Then we build the clone model and obtain initial clone item embedding vectors and user embedding vectors. We use a weighted sum to fuse the auxiliary item embeddings with the clone item embeddings to obtain fused item embeddings. The fused item embeddings and user embeddings are further used to calculate the predicted rating of the user for the item. The calculation of the weights for the auxiliary item embeddings and clone item embeddings is combined with the attention mechanism. For a $(user, item)$ interaction pair, we use a trainable neural network layer to calculate the attention coefficients for the auxiliary item embeddings and clone item embeddings of the interaction pair. Finally, using the same training mode as PTD, we can implement a new attack method called \textbf{PTA} (\textbf{P}artial \textbf{T}arget data and \textbf{A}uxiliary data).

Third, when the attacker can perform limited queries on the target model, we design the stealing function to extract the recommendation list obtained by querying the target model. Inspired by the previous work~\cite{yue2021black}, we divide the information contained in the recommendation list into two categories: recommended item information and ranking information. Recommended item information implies that the target model considers the user as more likely to like the recommended item than other items. 
We randomly sample some items from other items, called negative items. During the training process, we encourage the clone model to provide higher predicted ratings for positive items than negative items when making recommendations for the current user. Ranking information represents the target model’s belief that items ranked higher in the recommendation list is more preferred by the user than those ranked lower. Therefore, we design corresponding losses to encourage the clone model to provide higher predicted ratings for items ranked higher in the recommendation list than those ranked lower when making recommendations for the user. We call the attack algorithm with \textbf{P}artial \textbf{T}arget data and \textbf{Q}uery permission as \textbf{PTQ}, and call PTQ with the \textbf{A}uxiliary data as \textbf{PTAQ}.

In response to model stealing attacks on recommender systems, we also explore a defense strategy. 
We attempt to mix in some popular items to the recommendation list to mislead the clone to mimic the target model. These popular items, while not necessarily liked by users, are generally not hated. While achieving adequate defensive performance, it degrades the recommendation performance of the target model, which is consistent with perturbation-based defenses in other fields~\cite{orekondy2019prediction, zhang2021adversarial}. 
Its drawback underscores the significance of our work and the need for more effective defense mechanisms.

To recap, the main contributions of this paper are:
\begin{itemize}[leftmargin = 10 pt]
    \item We outline model stealing attacks against recommender systems with various types of knowledge and propose multiple effective attack strategies accordingly.
    \item Through an attention mechanism, we fuse "cheap" auxiliary data and "precious" target data. Besides, we design the stealing function to extract recommendation lists. These moves could help the clone model to better approximate the target model.
    \item We validate that our attacks can pose a serious threat to the model privacy of recommender systems on multiple datasets and various recommender systems. Even if the target model incorporates defense mechanisms, our algorithms still have good attack performance.
\end{itemize}

\section{Related Work}
\textbf{Recommender system}~\cite{shani2011evaluating, pazzani2007content, ansari2000internet} models user preferences and item attributes to recommend a list of items that users are more likely to be interested in.
Most recommender systems~\cite{hu2008collaborative, mnih2007probabilistic} generate a unique representation vector for each user to model their preferences. 
For example, WRMF~\cite{hu2008collaborative} and PMF~\cite{mnih2007probabilistic} use matrix factorization to decompose the user rating matrix into two matrices: user embeddings and item embeddings. 
The likelihood of a given interaction can be represented as the inner product of the user embedding and the item embedding.

Recently, some studies~\cite{lam2006you, zhang2021graph, zhang2021membership, zhang2022comprehensive} have pointed out that recommender systems face serious data privacy risks. For example, attackers can detect whether a user has engaged in the training process of the target model by using auxiliary data and query permission~\cite{yuan2023interaction, wang2022debiasing, zhang2021membership}. Such attacks are called membership inference attacks~\cite{shokri2017membership, rahman2018membership, hu2022membership}. 
The research on the model privacy of recommender systems is, however, still in its early stages. 
Countermeasures~\cite{zhan2010privacy, aimeur2008alambic,  mcsherry2009differentially} against privacy attacks on recommender systems have also been proposed. 
For instance, Zhang et al.~\cite{zhang2021membership} proposes to use randomized recommendation lists to resist membership inference attacks on recommender systems. 

\noindent\textbf{Model stealing attack}~\cite{miura2021megex, takemura2020model, tramer2016stealing, kleinbaum2002logistic} aims to steal internal information of the target model, including hyperparameters~\cite{wang2018stealing}, architecture~\cite{oh2019towards}, etc. Model stealing attacks can also be used to realize functional stealing attacks~\cite{jagielski2020high, orekondy2019knockoff, kleinbaum2002logistic}, which means building a clone model to imitate the predictions of the target model. 
The fine-tuned clone model can replace the target model in some ways. In addition, the clone model can be used for subsequent adversarial attacks~\cite{madry2017towards, finlayson2019adversarial}, membership inference attacks~\cite{shokri2017membership, rahman2018membership}, etc.  
Model stealing attacks have been widely studied in fields such as images~\cite{wang2022black, yu2020cloudleak} and graphs~\cite{wu2022model, he2021stealing}, while received little attention in recommender systems.
Yue et al.~\cite{yue2021black} proposed to utilize the autoregressive nature of sequential recommender system to steal its internal information. However, their method is only applicable to sequential recommender systems and lacks systematic evaluation of the performance of model stealing attacks. 

\section{Problem Formulation}
\subsection{Target Model}
After modeling users and items, the majority of recommender systems, such as matrix factorization-based recommenders~\cite{mnih2007probabilistic, rendle2012bpr}, assign a fixed-length embedding to each user and item and then use inner product operation to calculate the predicted rating of each user for each item. The following formula defines the predicted rating of user $i$ to item $j$.
The higher the predicted rating $r_{ij}$, the higher the likelihood of interaction between the two. 
\begin{equation}
r_{ij} =  \boldsymbol{p}_i \cdot \boldsymbol{q}_j,
\end{equation}
where $\boldsymbol{p}_i$ and $\boldsymbol{q}_j$ are embeddings of user $i$ and item $j$, respectively.

After calculating the predicted rating of the user $i$ for all items, the recommender system sorts these items based on their rating and recommends the highest-rated items to the user, excluding those that have already been interacted with.
\begin{equation}
\mathcal{R}_{i} = \mathop{\arg\max}\limits_{j \notin \mathcal{I}_{i}} {r_{ij}}.
\end{equation}
$\mathcal{I}_{i}$ represents the set of items that user $i$ has interacted with, and $\mathcal{R}_{i}$  the recommended item list for user $i$.
The notations described in this paper are summarized in an easy-to-read format in Table~\ref{notation}.

\begin{table}[]
\centering
\caption{Summary of the notations}
\begin{tabular}{l|l}
\hline
Notation                 & Description               \\ \hline
$\boldsymbol{p}_i$       & User $i$'s embedding                     \\
$\boldsymbol{q}_j$       & Item $j$'s embedding                     \\
$r_{ij}$                 & Predicted rating of user $i$ to item $j$ \\
$\mathcal{I}_{i}$       & Item set that user $i$ has interacted with \\
$\mathcal{R}_{i}^{target}$ & Target model's recommendation for user $i$ \\
$\mathcal{R}_{i}^{clone}$ & Clone model's recommendation for user $i$   \\
$\boldsymbol{q}^a$            & Auxiliary item embedding \\
$\boldsymbol{q}^c$            & Clone item embedding   \\
$\alpha'$     & Attention coefficient for $\boldsymbol{q}^c$ \\
$\beta'$      & Attention coefficient for $\boldsymbol{q}^a$ \\
$L_S$        & Stealing loss function \\
$L_r,L_p$        & Ranking loss and positive item loss  \\
m                     & The value of margin                                 \\
\hline
\end{tabular}
\label{notation}
\end{table}
\subsection{Threat Model}
\label{threatmodel}
\noindent\textbf{Adversary’s Goal.}
Model stealing attacks, also known as MSA, are presented in order to set up a local replica, also known as a clone model, of the target model.  
The fidelity, which promotes the clone model to deliver the same prediction for each sample as the target model, is the evaluation metric used for most model stealing attacks. In recommender systems, the target model tries to generate a user-preferred item list for each user. Thus, we hope that the clone model will recommend items for the same user that align with the recommendations made by the target model. We use Agreement (Agr)~\cite{yue2021black} to analyze the attack performance of model stealing attacks on recommender systems. The Agr for user $i$ is constructed according to the following:
\begin{equation}
{Agr}_{i} = \frac{|\mathcal{R}^{target}_{i} \cap \mathcal{R}^{clone}_{i}|}{|\mathcal{R}_{i}^{target}|},
\end{equation}
where $\mathcal{R}^{target}_{i}$ and $\mathcal{R}^{clone}_{i}$ are recommendations of the target model and the clone model for user $i$, respectively.
$|\mathcal{R}_{i}^{target}|$ is the length of recommendation, which equals to $|\mathcal{R}_{i}^{clone}|$.

\noindent\textbf{Adversary’s Knowledge.}
In this part, we reviewed three forms of knowledge that an attacker could have. 
Firstly, compared with previous works~\cite{li2016data, wu2021triple} that use all the target data, we assume that the attacker is able to acquire a minute portion (10\% in our experiments) of the target data. 
This data may originate from the target model's public dataset or accounts stolen by the attacker. Secondly, in reference to previous privacy attacks~\cite{zhang2021membership, wang2022debiasing}, we suppose that the adversary may have an auxiliary dataset that comes from the same distribution as the target data. In this assumption, the attacker may be a commercial competitor of the target model, with auxiliary data similar to the target data in terms of item sets. Finally, the attacker may obtain the recommendations of the target model on the available target data by query permission like QSD~\cite{yue2021black}.

\begin{table}[]
\centering
\caption{Attack knowledge of our methods}
\begin{tabular}{l|c|c|c}
\hline
Method      & \textbf{P}artial \textbf{T}arget data & \textbf{A}uxiliary data & \textbf{Q}uery \\
\hline
PTA          &  \checkmark   &  \checkmark  &                 \\
PTQ          &  \checkmark   &              &  \checkmark     \\
PTAQ         &  \checkmark   &  \checkmark  &  \checkmark     \\
\hline
\end{tabular}
\label{knowledge2}
\vspace{-0.3cm}
\end{table}
\section{Attack methodology}
In this section, we propose three algorithms based on the attacker's permissions. 
The methods and corresponding knowledge required are shown in Table~\ref{knowledge2}.
We will demonstrate step-by-step how to construct a clone model that imitates the target model's recommendation by using the three types of attack knowledge. 

In general, when the attacker can obtain a portion of the target data, we force the clone model to provide high ratings for interacted items(Section \ref{PTD}). When the attacker can obtain auxiliary data, we fuse the auxiliary item embeddings obtained from the auxiliary data into the item embeddings of the clone model and use an attention mechanism to assign reasonable weights to them (Section \ref{sec: PTA}). Finally, for attackers who have query permission to the target model, we design a stealing function to extract two kinds of information from the target model, namely ranking information and recommended item information (Section \ref{sec: query}).

\subsection{Partial Target Data}
\label{PTD}
In this subsection, we introduced how to establish a clone model in the scenario where the attacker only has access to partial target data. 
The training objective of the target model is to furnish personalized recommendations to the target users. Therefore, during the training phase, the target model is inclined to assign higher ratings to items interacting with these users. 
Leveraging this characteristic, we employ the same training target to train the clone model, with the expectation that it will also provide personalized recommendations to available target users. 

Consider, for instance, the Bayesian Personalized Ranking (BPR) model. The input format of the BPR model is a triplet of the type (user ID, positive item ID, negative items ID).
Specifically, the user ID and positive item ID come from the user’s interaction record, while the negative items are sampled from the remaining items. For instance, (user ID 5, positive item ID 4, negative item ID 3) indicates that user 5 has interacted with item 4, not item 3. When training the recommender systems, BPR establishes a pair-wise loss function for each pair of positive and negative items. The goal of this function is to produce higher ratings for positive items compared to the ratings of negative items. The following gives BPR's loss function:
\begin{equation}
    L_{BPR}(r_{ip}, r_{in}) = -\ln \sigma(r_{ip} - r_{in}).
    \label{BPR}
\end{equation}
$r_{ip}$ and $r_{in}$ are user $i$'s predicted ratings for positive item $p$ and negative item $n$.
$\sigma(\cdot)$ is the sigmoid function.

\subsection{Auxiliary Data}
\label{sec: PTA}
\begin{figure*}[h]
  \centering
  \includegraphics[width=0.8\linewidth]{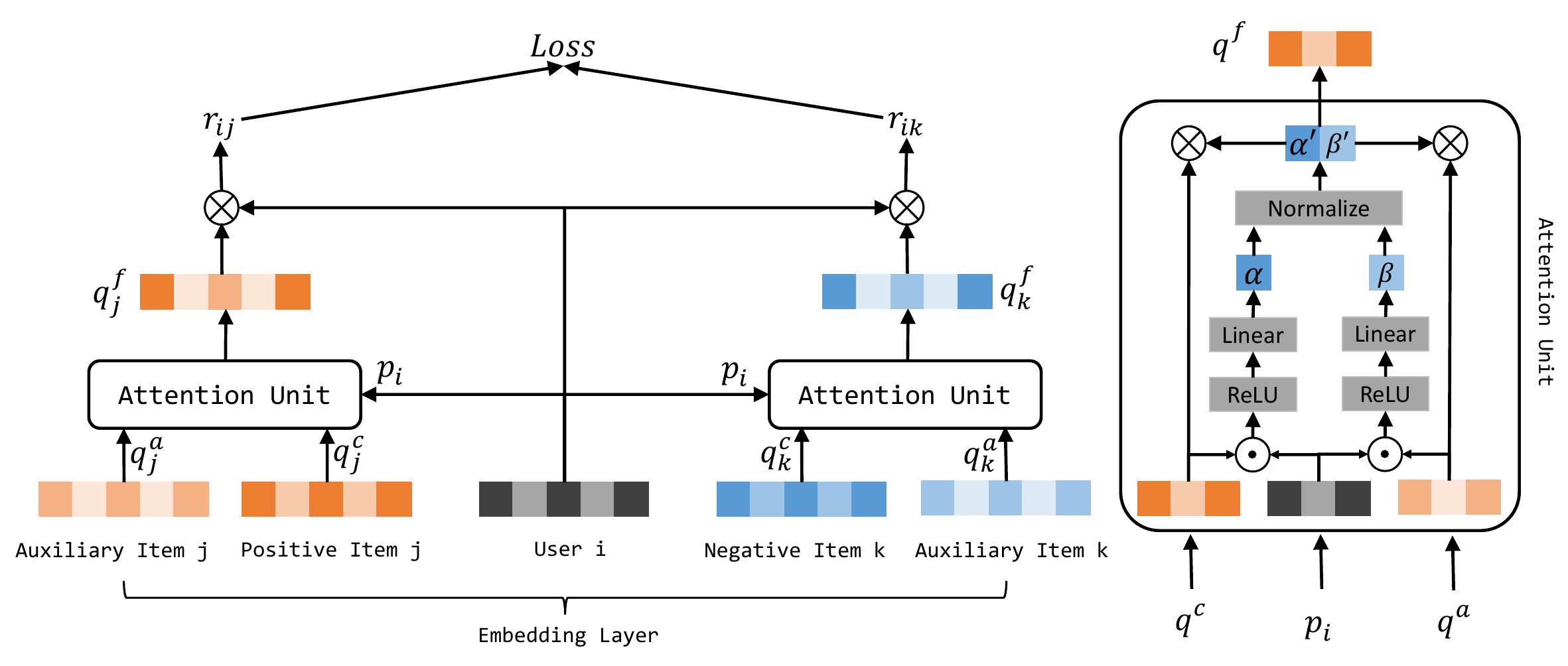}
  \caption{The process of utilizing the auxiliary data.}
  \label{auxdata}
\end{figure*}

In this subsection, we will discuss how incorporating auxiliary data into our approach will work. This kind of scenario often takes place when the attacker is a rival of the target model or when the attacker has obtained auxiliary data from other open platforms. Considering that the auxiliary data is independent of the target model, simply employing the auxiliary data to train the clone model, as Section~\ref{PTD} does, does not assist the clone model in becoming a precise imitation of the target model. 
Undeniably, auxiliary data and target data are often generated by similar users, which may make them contain similar behavior patterns. Then, fusing information of the auxiliary data into the modeling of the clone model's item attribute is likely to better help to emulate the target model.

First, in order to mine the hidden information in the auxiliary data, we train an auxiliary model on it (note that the auxiliary model is not the clone model). The item embeddings of the auxiliary data trained on the auxiliary model with the same architecture as the clone model has the item embeddings aligned with the ones in the clone model. That is, each row in the item embedding matrix of the clone model corresponds to each row in the item embedding matrix of the auxiliary model. Then, we employ a weighted addition approach to combine the clone item embeddings and auxiliary item embeddings, utilizing an attention mechanism. This mechanism is a unique structure in machine learning models that facilitates automatic learning and calculation of the respective contributions of the clone and auxiliary item embeddings towards the ultimate fused item embedding. 

Our approach involves the implementation of a single-layer neural network that utilizes a non-linear activation function to compute the weight of both the clone item embedding and auxiliary item embedding. 
Furthermore, in order to tailor the weights for individual users, we incorporate user embeddings into the weight computation procedure.
To predict user $i$'s rating of item $j$, $r_{ij}$, we first obtain user embedding $\boldsymbol{p}_i$, clone item embedding $\boldsymbol{q}_j^c$, and auxiliary item embedding $\boldsymbol{q}_j^a$. 
Then, the attention coefficient between user embedding and clone item embedding, auxiliary item embedding is calculated by the following formula.

\begin{algorithm}[htbp]
    \caption{Algorithm of PTAQ}
    \label{alg} 
    \KwIn{Available target data $D_t$; the auxiliary data $D_a$;  Query permission to the target model $M_{target}$;}
    \KwOut{The trained clone model $M_{clone}$} 
    Train $\boldsymbol{p}^a, \boldsymbol{q}^a$ with $D_a$ \tcp*{Parameters of $M_{aux}$}
    Random Initialize $\boldsymbol{p},\boldsymbol{q}^c$ \tcp*{Parameters of $M_{clone}$}
    \While(\tcp*[f]{train $M_{clone}$ with $D_t$}){not converge}{ 
        $Loss\leftarrow 0$\;
        \For(\tcp*[f]{user-item pair}){($i$, $j$) $\in D_t$} {
            $r_{ij}\leftarrow \boldsymbol{p}_i\cdot$weighted\_sum$(\boldsymbol{q}_j^c, \boldsymbol{q}_j^a)$  \tcp*{Eq.~\ref{att4}}
            \For{$k \in$ Sampled Negative Items}{
                $r_{ik}\leftarrow \boldsymbol{p}_i\cdot$weighted\_sum$(\boldsymbol{q}_k^c, \boldsymbol{q}_k^a)$  \tcp*{Eq.~\ref{att4}}
                $Loss += L_{BPR}(r_{ij}, r_{ik})$ \tcp*{Eq.~\ref{BPR}}
            }
        }
        Update $\boldsymbol{p},\boldsymbol{q}^c$ by $Loss$\;
    }
    \While(\tcp*[f]{Fine-tune $M_{clone}$}){not converge}{
        $Loss\leftarrow 0$\;
        \For{user $i \in D_t$} {
            $R^{target}_i\leftarrow$ $M_{target}$'s recommendation list for $i$\;
            \For{$j \in R^{target}_i$} {
                $r_{ij}\leftarrow \boldsymbol{p}_i\cdot$weighted\_sum$(\boldsymbol{q}_j^c, \boldsymbol{q}_j^a)$  \tcp*{Eq.~\ref{att4}}
                \For{$k \in$ Sampled Negative Items}{
                    $r_{ik}\leftarrow \boldsymbol{p}_i\cdot$weighted\_sum$(\boldsymbol{q}_k^c, \boldsymbol{q}_k^a)$  \tcp*{Eq.~\ref{att4}}
                    $Loss += L_{Hinge}(r_{ij}, r_{ik})$ \tcp*{Eq.~\ref{Hinge}}
                }
                $j'\leftarrow $ the next item of $j$ in $R_i^{target}$\;
                $r_{ij'}\leftarrow \boldsymbol{p}_i\cdot$weighted\_sum$(\boldsymbol{q}_{j'}^c, \boldsymbol{q}_{j'}^a)$  \tcp*{Eq.~\ref{att4}}
                $Loss += L_{BPR}(r_{ij}, r_{ij'})$ \tcp*{Eq.~\ref{BPR}}
            }
        }
        Update $\boldsymbol{p},\boldsymbol{q}^c$ by $Loss$\;
    }
    
\end{algorithm}

\begin{equation}
    \alpha = \boldsymbol{w}^T ReLU (\boldsymbol{p}_i \odot \boldsymbol{q}_j^c) + b,\quad \beta = \boldsymbol{w}^T ReLU (\boldsymbol{p}_i \odot \boldsymbol{q}_j^a) + b,
    \label{att1}
\end{equation}
where $\odot$ represents Hadamard product(element-wise product). $\boldsymbol{w}$ and $b$ are parameters of the neural network. ReLU is a non-linear activation function.
In order to keep the magnitude of the final item embedding constant, we normalize the obtained $\alpha$ and $\beta$, to guarantee that the sum of $\alpha'$ and $\beta'$ equals 1.
\begin{equation}
    \alpha' = \frac{e^{\alpha}}{e^{\alpha} + e^{\beta}}, \quad\beta' = \frac{e^{\beta}}{e^{\alpha} + e^{\beta}}.
    \label{att2}
\end{equation}
After calculating the weight pairs, we fuse the clone and the auxiliary item embedding to obtain the fused item embedding. 
\begin{equation}
    \boldsymbol{q}^f_j = \alpha' \boldsymbol{q}^c_j + \beta' \boldsymbol{q}^a_j.
    \label{att3}
\end{equation}
We perform inner product operations on the user embeddings and fused item embeddings to obtain their predicted ratings.
\begin{equation}
    r_{ij} =  \boldsymbol{p}_i \cdot \boldsymbol{q}_j^f.
    \label{att4}
\end{equation}
The process of utilizing the auxiliary data is shown in Fig~\ref{auxdata}. 

\subsection{Query Permission}
\label{sec: query}
In this subsection, we assume that the attacker has query permission and can obtain the ordered recommendation list of the target data from the target model. 
Inspired by the previous work~\cite{yue2021black}, we extract two pieces of information from the recommendation, namely recommended item information and ranking information. The recommended item information relates to the fact that users are thought to interact with the recommended items more likely than with other items.
The ranking information indicates that the user is more likely to prefer the items at the top of the recommendation list rather than those at the bottom. 
We adopt the margin hinge loss to extract the recommended item information.
\begin{equation}
    L_{Hinge}(r_{ip}, r_{in}) = max(0, m - (r_{ip} - r_{in})),
    \label{Hinge}
\end{equation}
where $r_{ip}$ and $r_{in}$ are user $i$'s predicted ratings for positive item $p$ and negative item $n$.
$m$ is the value of the margin.
The margin hinge loss instructs the model to adjust its parameters in such a way that the difference in rating between the positive and negative items is greater than the margin value. This explicit design allows us to better control the constraints on the model's predicted ratings. 
We use the BPR loss function in Section~\ref{PTD} to utilize the ranking information. 
We designed a stealing loss function, which considered both ranking information and recommended item information. 
\begin{equation}
L_S^i = L_r^i + L_p^i,
\end{equation}
where $L_S^i$, $L_r^i$, and $L_p^i$ are user $i$'s stealing loss, ranking loss, and positive item loss, respectively.
$L_r^i$ and $L_p^i$ are shown as follows:
\begin{equation}
L_r^i = \sum_{j \in \mathcal{R}^{target}_{i}} {L_{BPR}(r_{ij}, r_{ij'})},
\end{equation}
\begin{equation}
L_p^i = \sum_{j \in \mathcal{R}^{target}_{i}} \sum_{k\in n_j} {L_{Hinge}(r_{ij}, r_{ik})},
\end{equation}
where $\mathcal{R}^{target}_{i}$ is the target model's recommendation for user $i$.
$j'$ is the item after item $j$ in the recommendation list, and $n_j$ is the negative item set for item $j$.

\subsection{Overall Workflow}
In this subsection, we describe the workflow of our methods using BPR as the target model. Here we mainly focus on PTAQ with three types of knowledge, and its algorithm flow is shown in Alg. \ref{alg}.
We first train the auxiliary model $M_{aux}$ with auxiliary data $D_a$ (line 1). The parameters of $M_{aux}$ are denoted as $\boldsymbol{p}^a, \boldsymbol{q}^a$, which refer to user and item embeddings. 
After that, we randomly initialize $\boldsymbol{p}, \boldsymbol{q}^c$, the parameters of $M_{clone}$ (line 2), and carefully train the parameters by BPR loss(line 3-13). During the training, the predicted rating for each user-item pair is mainly based on Eq.\ref{att4}, which incorporates the weighted sum of item embedding of $M_{clone}$ and $M_{aux}$.
Once the converged $M_{clone}$ is obtained, $D_t$ is utilized again to fine-tune $M_{clone}$ (line 14-30). In the fine-tuning process, we iterate each user in $D_t$ and query $M_{target}$ with the user id to obtain $M_{target}$'s recommendation list $R_i^{target}$. After that, Hinge Loss and BPR Loss are adopted to extract positive item information and ranking information.

For PTA that lacks query permission compared to PTAQ, the fine-tuning phase (Line 14-30 of Alg. \ref{alg}) will be lacking. For PTQ that lacks auxiliary data, the function $weighted\_sum(q_i^c,q_i^a)$ ($i=j, k, or j'$ in Alg. \ref{alg}) will be replaced by the item embedding $q_i^c$ ($i=j, k, or j'$ in Alg. \ref{alg}) of the clone model.

\section{Evaluation}
\label{Evaluation}
In this section, we evaluate model stealing attacks against multiple classic recommender systems on diverse datasets.
We assess the attack methods on three real-world datasets, including ML-1M(MovieLens-1M)~\cite{harper2015movielens}, Ta-feng\footnote{https://www.kaggle.com/datasets/chiranjivdas09/ta-feng-grocery-dataset}, and Steam\footnote{https://www.kaggle.com/datasets/tamber/steam-video-games}. We randomly split each dataset evenly into two disjoint subsets: a target dataset and an auxiliary dataset,  and report the average Agreement (Agr)~\cite{yue2021black} of 5 runs for each method to avoid bias.
The target and auxiliary data do not have a common user, but they do have common items.
We conduct experiments against three classic recommender systems, namely BPR(Bayesian Personalized Ranking), NCF(Neural Collaborative Filtering), and LMF(Logistic Matrix Factorization).
We adopt PTD~\cite{li2016data, nguyen2023poisoning} and QSD~\cite{yue2021black} as baselines.
The detailed experimental setup is included in the Appendix~\ref{Setup}.

Following the experimental setup, we investigate the performance of different attack methods across multiple datasets in Section~\ref{Attack Performance on Various Datasets} and carry out an in-depth analysis of the impact of attack knowledge in Section~\ref{Attack Performance with Different Knowledge}.
Furthermore, in Section~\ref{Aux details} we go through the auxiliary data in detail.
In addition, we discuss the influence of query budget and stealing loss function in the Appendix~\ref{appendix}.

\begin{table*}[]
\centering
\caption{Attack performance(Agr) on various datasets.}
\begin{tabular}{l|ccc|ccc|ccc}
\hline\noalign{\smallskip}
\multirow{2}{*}{Algorithms} & \multicolumn{3}{c|}{ML-1M} & \multicolumn{3}{c|}{Ta-feng}    & \multicolumn{3}{c}{Steam}  \\ \noalign{\smallskip}\cline{2-10} \noalign{\smallskip}
                           & BPR     & NCF    & LMF    & BPR     & NCF     & LMF     & BPR     & NCF    & LMF    \\
\noalign{\smallskip}\hline\noalign{\smallskip}
QSD                        & 0.0533  & 0.2727 & 0.0487 & 0.0598  & 0.4398  & 0.2371  & 0.3561  & 0.8710 & 0.3721 \\
PTD                        & 0.4987  & 0.3633 & 0.2747 & 0.4718  & 0.6526  & 0.4537  & 0.7773  & 0.8512 & 0.0181 \\
\noalign{\smallskip}\hline\noalign{\smallskip}
PTQ                        & 0.5107  & 0.3633 & 0.2873 & 0.4845  & 0.7518  & 0.7986  & 0.9276  & 0.9813 & 0.2496 \\
PTA                        & 0.6280  & 0.4400 & 0.5266 & 0.5135  & 0.6545  & 0.5488  & 0.8957  & 0.8899 & 0.6817 \\
PTAQ                       & \textbf{0.6506}  & \textbf{0.4407} & \textbf{0.5627} & \textbf{0.5141}  & \textbf{0.7692}  & \textbf{0.8038}  & \textbf{0.9476}  & \textbf{0.9833} & \textbf{0.9553} \\
\noalign{\smallskip}\hline
\end{tabular}
\label{Main Results}
\end{table*}

\begin{figure*}[h]
  \centering
  \includegraphics[width=0.8\linewidth]{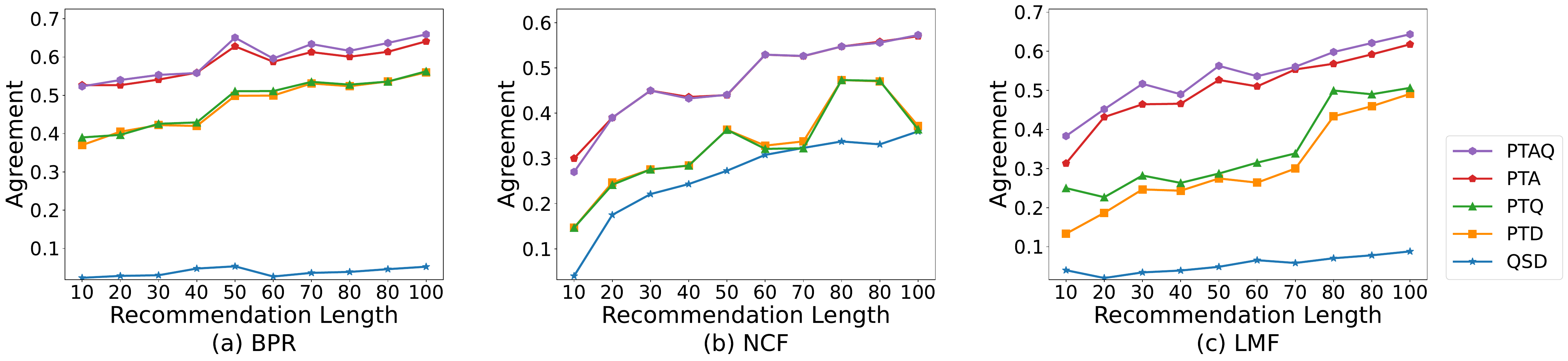}
  \caption{Attack performance with various recommendation lengths on ML-1M.}
  \label{cutoff}
\end{figure*}

\begin{figure*}[h]
  \centering
  \includegraphics[width=0.8\linewidth]{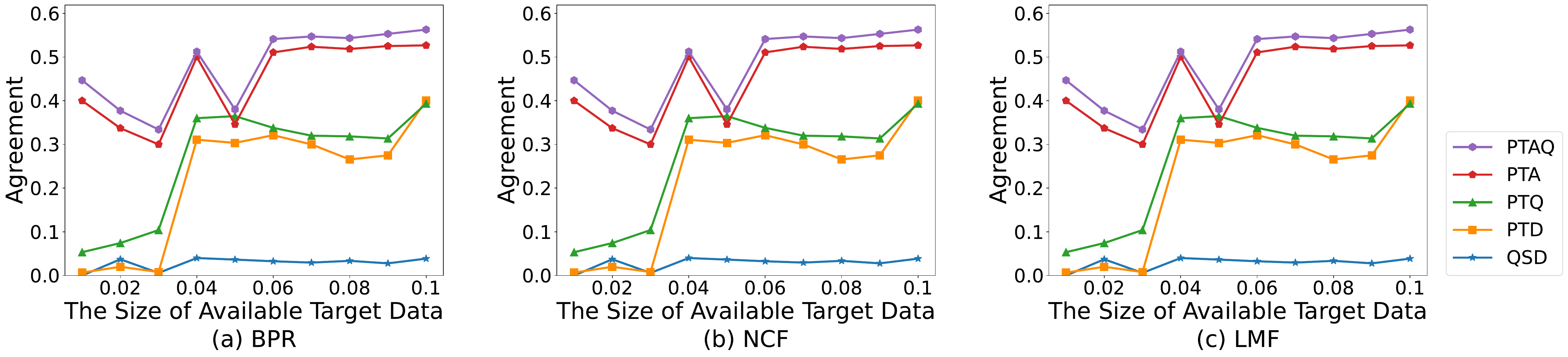}
  \caption{Attack performance with different sizes of available target data on ML-1M.}
  \label{init rate}
\end{figure*}

\subsection{Attack Performance on Various Datasets}
\label{Attack Performance on Various Datasets}
In this section, we perform experiments on ML-1M, Ta-feng, and Steam datasets against three recommendation models, including BPR, NCF, and LMF. 
The experimental results show that all these methods can steal model information to some extent. Based on the experimental results in Table~\ref{Main Results}, we draw the following conclusions:
\begin{itemize}[leftmargin = 10 pt]
\item Auxiliary data and query permission can effectively improve the performance of model stealing attacks. PTAQ, which fuses both auxiliary data and query feedback information, achieves the best attack performance in all scenarios. 
\item The QSD algorithm has low performance across several datasets. 
QSD was designed for sequential recommender systems, and its performance is dependent on producing a large number of fake queries, which are not applicable to our assumptions. 
In the appendix, we compare the performance of our methods with QSD on sequential recommendation models.
\item It is difficult to steal the model information of LMF when we can only obtain a few target data. On the ML-1M, Ta-feng, and Steam datasets, PTD’s stealing performance on LMF was only 27.47\%, 45.37\%, and 1.81\%, respectively. 
When the attacker makes use of ample external information, such as auxiliary data and query feedback, this issue is significantly reduced.
\end{itemize}


\subsection{Performance with Different Knowledge}
\label{Attack Performance with Different Knowledge}

\subsubsection{Recommendation Length.}
In this part, we set the recommendation list length from 10 to 100 and analyze the impact of the recommendation list length on the performance of the attack method in Fig~\ref{cutoff}. Typically, the effectiveness of algorithms tends to increase with the lengthening of recommendation lists. On NCF, PTA is comparable to that of PTAQ, and PTD is comparable to that of PTQ, which is consistent with the experimental results in Section~\ref{Main Results}. It is probably because of the large number of user interactions that are recorded in the ML-1M dataset. As a consequence of this, the available target data and auxiliary data contain almost as much additional information as the recommendation lists can provide. 
While it can be seen from Table~\ref{Main Results} that PTQ is significantly improved when compared to PTD in the sparsely interacting \textbf{Steam} dataset.
\begin{figure*}[h]
  \centering
  \includegraphics[width=0.8\linewidth]{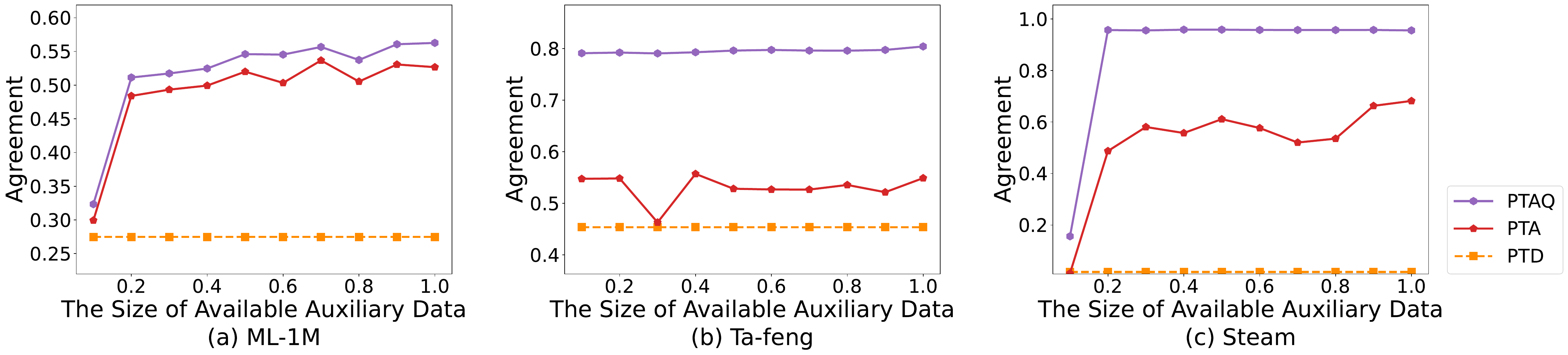}
  \caption{Attack performance with different sizes of available auxiliary data aginst LMF.}
  \label{aux_rate}
\end{figure*}

\begin{figure*}[h]
  \centering
  \includegraphics[width=0.8\linewidth]{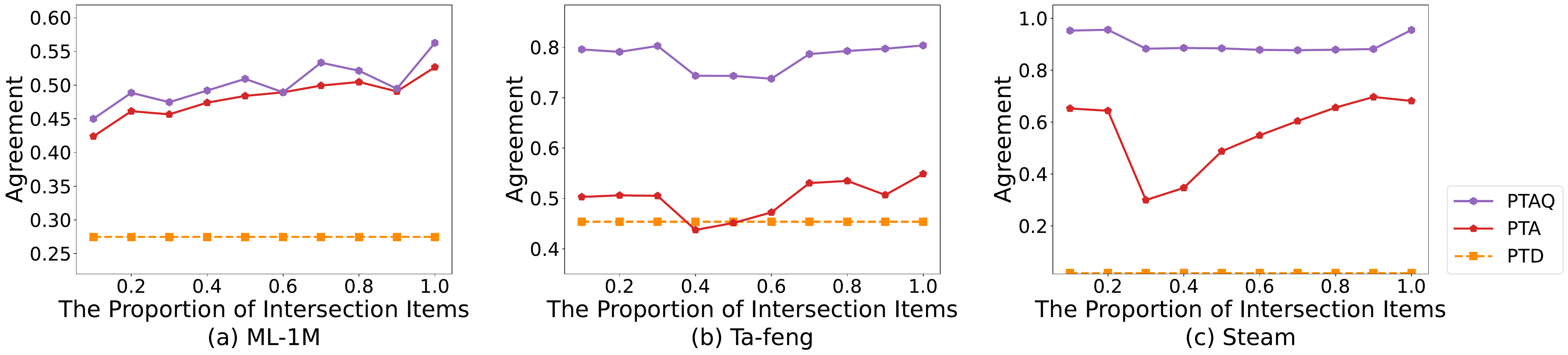}
  \caption{Attack performance with different proportions of intersection items aginst LMF.}
  \label{coin_rate}
\end{figure*}

\begin{table*}[]
\centering
\caption{Attack performance with different ways to exploit the auxiliary data}
\begin{tabular}{l|ccc|ccc|ccc}
\hline\noalign{\smallskip}
\multirow{2}{*}{Algorithms} & \multicolumn{3}{c|}{ML-1M} & \multicolumn{3}{c|}{Ta-feng}    & \multicolumn{3}{c}{Steam}  \\ \noalign{\smallskip}\cline{2-10} \noalign{\smallskip}
                           & BPR    & NCF    & LMF & BPR    & NCF    & LMF & BPR    & NCF    & LMF \\
\noalign{\smallskip}\hline\noalign{\smallskip}
PTA(Pre)                   & 0.4900 & 0.3793 & 0.2807     & 0.4629 & 0.6533 & 0.4598     & 0.7918 & 0.8594 & 0.0167     \\
PTAQ(Pre)                  & 0.4833 & 0.3793 & 0.2900     & 0.4598 & 0.7492 & 0.7886     & 0.9083 & 0.9841 & 0.1370     \\
PTA                        & 0.6280 & 0.4400 & 0.5266     & 0.5135 & 0.6545 & 0.5488     & 0.8957 & 0.8899 & 0.6817     \\
PTAQ                       & \textbf{0.6506} & \textbf{0.4407} & \textbf{0.5627}     & \textbf{0.5141} & \textbf{0.7692} & \textbf{0.8038}     & \textbf{0.9476} & \textbf{0.9833} & \textbf{0.9553}     \\
\noalign{\smallskip}\hline\noalign{\smallskip}
\end{tabular}
\label{wuad}
\end{table*}

\subsubsection{Sizes of Available Target Data.} 
Figure~\ref{init rate} analyzed attack performances with respect to the varying size of the target data accessible to the attacker.  
We can obeserve that the efficacy of the attack method is proportional to the magnitude of the available target dataset. The main reason is that more available target data can provide us with more interactions and query feedback information, thereby improving the attack effect of the clone model. 

\subsubsection{Knowable Model Architecture}
\begin{figure*}[h]
  \centering
  \includegraphics[width=0.8\linewidth]{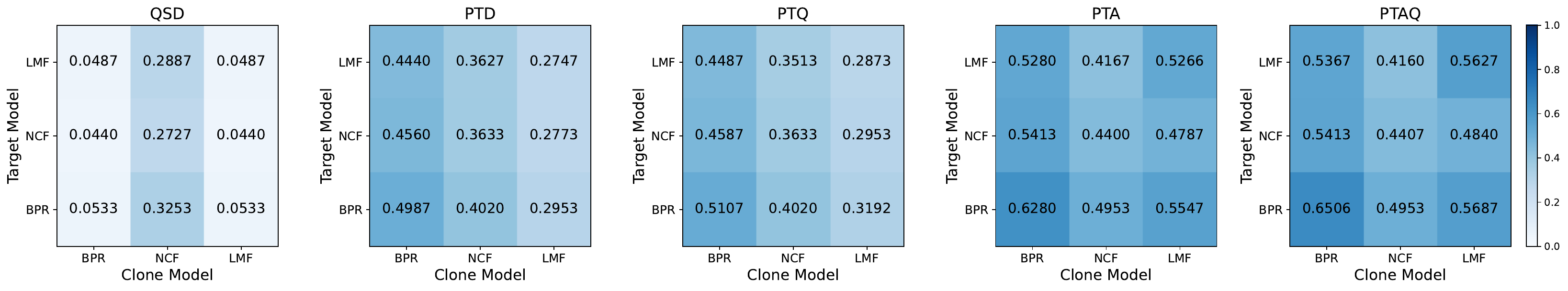}
  \caption{Attack performance with different model architectures on ML-1M.}
  \label{unware}
\end{figure*}

. In this section, we eliminate the assumption that the clone model exhibits identical architecture to that of the target model.
Figure~\ref{unware} depicts the utilization of heatmaps to visually represent the attack efficacy across various permutations of the target model and the clone model's architectures.
The vertical axis denotes the structural design of the target model, while the horizontal axis represents the structural design of the clone model. The visual representation of stronger attack effects is denoted by darker colors. 
Based on the empirical evidence presented in Figure~\ref{unware}, the following inferences can be made: 
\begin{itemize}[leftmargin = 10 pt]
    \item In scenarios where the attacker lacks knowledge of the underlying architecture of the target model, our attack capabilities remain adequate. 
    \item The utilization of BPR as the clone model shows superior attack efficacy in comparison to alternative models. 
    This finding suggests that prior knowledge of the target model's architecture is not a prerequisite. Identifying an appropriate architecture for the clone model is possible to enhance the attack performance. 
\end{itemize}

\begin{figure*}[h]
  \centering
  \includegraphics[width=0.8\linewidth]{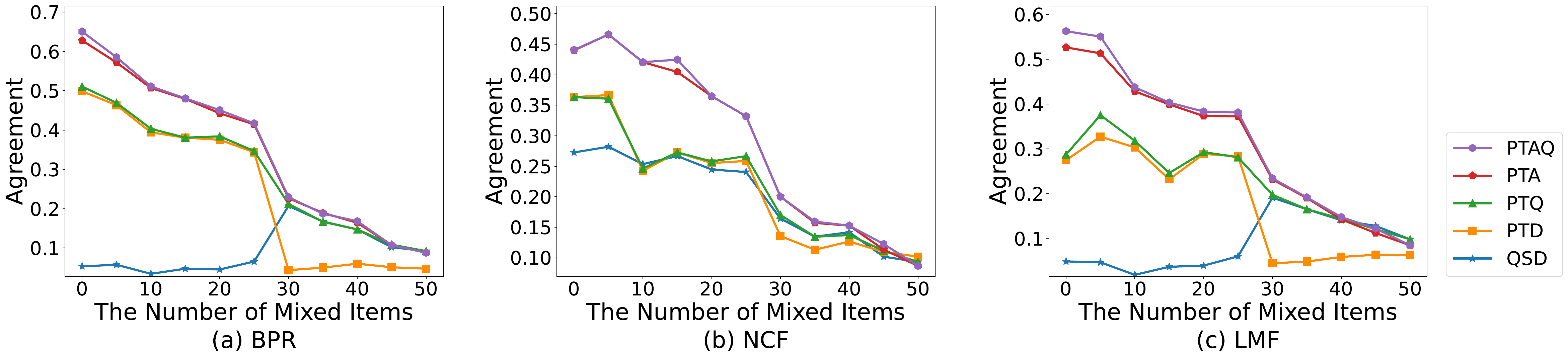}
  \caption{Attack performance with mixed recommendations on ML-1M.}
  \label{def}
\end{figure*}
\begin{figure}[h]
  \centering
  \includegraphics[width=0.7\linewidth]{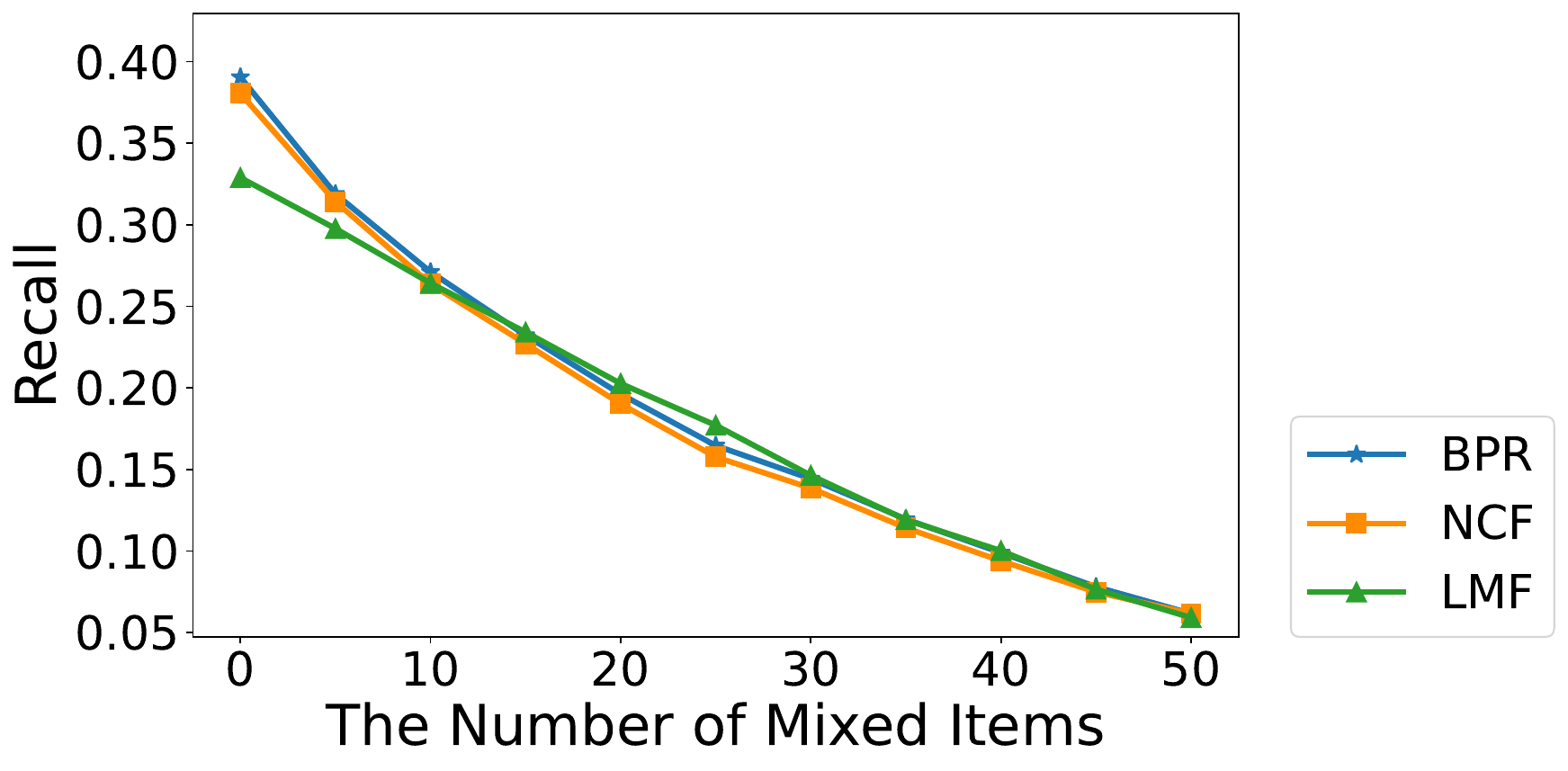}
  \caption{Recommendation performance with mixed recommendations on ML-1M.}
  \label{def2}
\end{figure}

\subsection{Auxiliary Dataset}
\label{Aux details}
\subsubsection{Attention Mechanism or Pretraining.}
\label{Ways to Utilize Auxiliary Dataset}
In this section, an alternative approach to exploit auxiliary data is investigated. 
The method called Pretraining(Pre) uses the auxiliary item embeddings as the initial state of the item embeddings in the clone model. We compare the performance of PTA and PTAQ using pretraining and attention mechanisms on three datasets in Table~\ref{wuad}.
It can be observed that using the attention mechanism to fuse auxiliary data can achieve more powerful attack performance than the pretraining method, regardless of whether the attacker has query permission.

\subsubsection{Sizes of Available Auxiliary Data.}
Figure~\ref{aux_rate} analyzed attack performances with respect to the varying size of available auxiliary data. 
From the experimental results, we find that even with limited auxiliary data, our method can produce a significant improvement in terms of PTD. Furthermore, the performance of PTA and PTAQ tends to be proportional to the number of auxiliary data.

\subsubsection{Sizes of Iteractions Items.}
We adjusted the overlap ratio(from 10\% to 100\%) between items in the auxiliary data and target data, meaning attackers can only use the auxiliary item embeddings overlapping with the target items for their modeling. The experimental results in Figure~\ref{coin_rate} show that, the utilisation of auxiliary data, despite \textbf{a low overlap ratio} with the target data, demonstrates a noticeable enhancement in the stealing performance.

\section{Defense}
\label{Defense}
It is of the utmost need to develop efficient techniques, as soon as possible, to reduce the harm caused by model stealing attacks. 
In this section, we construct a defensive method based on altering the recommendation list of the target model.
We introduce inaccuracies into the query feedback information that the attacker obtained in order to mislead them into mimicking recommendation behaviors that are incompatible with the target model.

Specifically, we first calculate item popularity based on the number of times it interacts with users. Then for each recommendation, we select some recommended items and replace them with popular items. 
While the inclusion of popular items may not necessarily align with the user's preferences, it is typically the case that such items do not elicit strong negative reactions from the user. 
We fix the length of the recommendation list to 50 and randomly select 5-25 items from the 100 most popular items to replace an equal number of items in the recommendation list. 
We use Recall~\cite{gunawardana2009survey} to assess the impact of the mixed recommendation on recommendation performance.
We report the changes in the attack performance of different algorithms and the recommendation performance of the target model in Fig~\ref{def} and Fig~\ref{def2}, respectively. Based on the experimental results, we draw the following conclusions: 
\begin{itemize}[leftmargin = 10 pt]
    \item The mixed defense strategy can effectively resist model stealing attacks. This is due to two factors. On the one hand, mixing popular items into the recommendation list misleads attackers utilizing query permission.
    On the other hand, adding some random popular items to the recommendation list pushes the target model to give recommendations based on not only user preferences but also randomness, making it more difficult for the cloned model to imitate the target model’s prediction behavior.
    The decreases in the attack performance of PTD and PTA also verify this assertion.  
    \item Although the mixing strategy can resist model stealing attacks to a certain extent, it inevitably causes a decrease in the recommendation performance of the target model. 
    As the number of mixed items increases, the defense performance gradually improves, but the recommendation performance continues to decline.
    We need to find an acceptable balance between the defense performance and the recommendation performance.
    It further illustrates the serious threat of model stealing attacks to recommender systems and the need to propose more robust and powerful defense strategies in the future.
\end{itemize}
\section{Conclusion and Future Work}
In this paper, we focus on model stealing attacks against recommender systems. 
We present multiple strategies to leverage the various types of knowledge that can be obtained by attackers.
We use an attention mechanism to fuse auxiliary data with target data and design a stealing function to extract the recommendation list of the target model. In future work, we will investigate how to better guard against model stealing attacks on recommender systems.

\clearpage
\bibliographystyle{ACM-Reference-Format}
\bibliography{ref}


\begin{thebibliography}{48}


\ifx \showCODEN    \undefined \def \showCODEN     #1{\unskip}     \fi
\ifx \showDOI      \undefined \def \showDOI       #1{#1}\fi
\ifx \showISBNx    \undefined \def \showISBNx     #1{\unskip}     \fi
\ifx \showISBNxiii \undefined \def \showISBNxiii  #1{\unskip}     \fi
\ifx \showISSN     \undefined \def \showISSN      #1{\unskip}     \fi
\ifx \showLCCN     \undefined \def \showLCCN      #1{\unskip}     \fi
\ifx \shownote     \undefined \def \shownote      #1{#1}          \fi
\ifx \showarticletitle \undefined \def \showarticletitle #1{#1}   \fi
\ifx \showURL      \undefined \def \showURL       {\relax}        \fi
\providecommand\bibfield[2]{#2}
\providecommand\bibinfo[2]{#2}
\providecommand\natexlab[1]{#1}
\providecommand\showeprint[2][]{arXiv:#2}

\bibitem[A{\"\i}meur et~al\mbox{.}(2008)]%
        {aimeur2008alambic}
\bibfield{author}{\bibinfo{person}{Esma A{\"\i}meur}, \bibinfo{person}{Gilles Brassard}, \bibinfo{person}{Jos{\'e}~M Fernandez}, {and} \bibinfo{person}{Flavien~Serge Mani~Onana}.} \bibinfo{year}{2008}\natexlab{}.
\newblock \showarticletitle{Alambic: a privacy-preserving recommender system for electronic commerce}.
\newblock \bibinfo{journal}{\emph{International Journal of Information Security}} \bibinfo{volume}{7}, \bibinfo{number}{5} (\bibinfo{year}{2008}), \bibinfo{pages}{307--334}.
\newblock


\bibitem[Ansari et~al\mbox{.}(2000)]%
        {ansari2000internet}
\bibfield{author}{\bibinfo{person}{Asim Ansari}, \bibinfo{person}{Skander Essegaier}, {and} \bibinfo{person}{Rajeev Kohli}.} \bibinfo{year}{2000}\natexlab{}.
\newblock \bibinfo{title}{Internet recommendation systems}.
\newblock
\newblock


\bibitem[Finlayson et~al\mbox{.}(2019)]%
        {finlayson2019adversarial}
\bibfield{author}{\bibinfo{person}{Samuel~G Finlayson}, \bibinfo{person}{John~D Bowers}, \bibinfo{person}{Joichi Ito}, \bibinfo{person}{Jonathan~L Zittrain}, \bibinfo{person}{Andrew~L Beam}, {and} \bibinfo{person}{Isaac~S Kohane}.} \bibinfo{year}{2019}\natexlab{}.
\newblock \showarticletitle{Adversarial attacks on medical machine learning}.
\newblock \bibinfo{journal}{\emph{Science}} \bibinfo{volume}{363}, \bibinfo{number}{6433} (\bibinfo{year}{2019}), \bibinfo{pages}{1287--1289}.
\newblock


\bibitem[Gunawardana and Shani(2009)]%
        {gunawardana2009survey}
\bibfield{author}{\bibinfo{person}{Asela Gunawardana} {and} \bibinfo{person}{Guy Shani}.} \bibinfo{year}{2009}\natexlab{}.
\newblock \showarticletitle{A survey of accuracy evaluation metrics of recommendation tasks.}
\newblock \bibinfo{journal}{\emph{Journal of Machine Learning Research}} \bibinfo{volume}{10}, \bibinfo{number}{12} (\bibinfo{year}{2009}).
\newblock


\bibitem[Harper and Konstan(2015)]%
        {harper2015movielens}
\bibfield{author}{\bibinfo{person}{F~Maxwell Harper} {and} \bibinfo{person}{Joseph~A Konstan}.} \bibinfo{year}{2015}\natexlab{}.
\newblock \showarticletitle{The movielens datasets: History and context}.
\newblock \bibinfo{journal}{\emph{Acm transactions on interactive intelligent systems (tiis)}} \bibinfo{volume}{5}, \bibinfo{number}{4} (\bibinfo{year}{2015}), \bibinfo{pages}{1--19}.
\newblock


\bibitem[He et~al\mbox{.}(2021)]%
        {he2021stealing}
\bibfield{author}{\bibinfo{person}{Xinlei He}, \bibinfo{person}{Jinyuan Jia}, \bibinfo{person}{Michael Backes}, \bibinfo{person}{Neil~Zhenqiang Gong}, {and} \bibinfo{person}{Yang Zhang}.} \bibinfo{year}{2021}\natexlab{}.
\newblock \showarticletitle{Stealing Links from Graph Neural Networks.}. In \bibinfo{booktitle}{\emph{USENIX Security Symposium}}. \bibinfo{pages}{2669--2686}.
\newblock


\bibitem[He et~al\mbox{.}(2017)]%
        {he2017neural}
\bibfield{author}{\bibinfo{person}{Xiangnan He}, \bibinfo{person}{Lizi Liao}, \bibinfo{person}{Hanwang Zhang}, \bibinfo{person}{Liqiang Nie}, \bibinfo{person}{Xia Hu}, {and} \bibinfo{person}{Tat-Seng Chua}.} \bibinfo{year}{2017}\natexlab{}.
\newblock \showarticletitle{Neural collaborative filtering}. In \bibinfo{booktitle}{\emph{Proceedings of the 26th international conference on world wide web}}. \bibinfo{pages}{173--182}.
\newblock


\bibitem[Hidasi et~al\mbox{.}(2015)]%
        {hidasi2015session}
\bibfield{author}{\bibinfo{person}{Bal{\'a}zs Hidasi}, \bibinfo{person}{Alexandros Karatzoglou}, \bibinfo{person}{Linas Baltrunas}, {and} \bibinfo{person}{Domonkos Tikk}.} \bibinfo{year}{2015}\natexlab{}.
\newblock \showarticletitle{Session-based recommendations with recurrent neural networks}.
\newblock \bibinfo{journal}{\emph{arXiv preprint arXiv:1511.06939}} (\bibinfo{year}{2015}).
\newblock


\bibitem[Hu et~al\mbox{.}(2022)]%
        {hu2022membership}
\bibfield{author}{\bibinfo{person}{Hongsheng Hu}, \bibinfo{person}{Zoran Salcic}, \bibinfo{person}{Lichao Sun}, \bibinfo{person}{Gillian Dobbie}, \bibinfo{person}{Philip~S Yu}, {and} \bibinfo{person}{Xuyun Zhang}.} \bibinfo{year}{2022}\natexlab{}.
\newblock \showarticletitle{Membership inference attacks on machine learning: A survey}.
\newblock \bibinfo{journal}{\emph{ACM Computing Surveys (CSUR)}} \bibinfo{volume}{54}, \bibinfo{number}{11s} (\bibinfo{year}{2022}), \bibinfo{pages}{1--37}.
\newblock


\bibitem[Hu et~al\mbox{.}(2008)]%
        {hu2008collaborative}
\bibfield{author}{\bibinfo{person}{Yifan Hu}, \bibinfo{person}{Yehuda Koren}, {and} \bibinfo{person}{Chris Volinsky}.} \bibinfo{year}{2008}\natexlab{}.
\newblock \showarticletitle{Collaborative filtering for implicit feedback datasets}. In \bibinfo{booktitle}{\emph{2008 Eighth IEEE international conference on data mining}}. Ieee, \bibinfo{pages}{263--272}.
\newblock


\bibitem[Jagielski et~al\mbox{.}(2020)]%
        {jagielski2020high}
\bibfield{author}{\bibinfo{person}{Matthew Jagielski}, \bibinfo{person}{Nicholas Carlini}, \bibinfo{person}{David Berthelot}, \bibinfo{person}{Alex Kurakin}, {and} \bibinfo{person}{Nicolas Papernot}.} \bibinfo{year}{2020}\natexlab{}.
\newblock \showarticletitle{High accuracy and high fidelity extraction of neural networks}. In \bibinfo{booktitle}{\emph{Proceedings of the 29th USENIX Conference on Security Symposium}}. \bibinfo{pages}{1345--1362}.
\newblock


\bibitem[Johnson(2014)]%
        {johnson2014logistic}
\bibfield{author}{\bibinfo{person}{Christopher~C Johnson}.} \bibinfo{year}{2014}\natexlab{}.
\newblock \showarticletitle{Logistic matrix factorization for implicit feedback data}.
\newblock \bibinfo{journal}{\emph{Advances in Neural Information Processing Systems}} \bibinfo{volume}{27}, \bibinfo{number}{78} (\bibinfo{year}{2014}), \bibinfo{pages}{1--9}.
\newblock


\bibitem[Kang and McAuley(2018)]%
        {kang2018self}
\bibfield{author}{\bibinfo{person}{Wang-Cheng Kang} {and} \bibinfo{person}{Julian McAuley}.} \bibinfo{year}{2018}\natexlab{}.
\newblock \showarticletitle{Self-attentive sequential recommendation}. In \bibinfo{booktitle}{\emph{2018 IEEE international conference on data mining (ICDM)}}. IEEE, \bibinfo{pages}{197--206}.
\newblock


\bibitem[Kleinbaum et~al\mbox{.}(2002)]%
        {kleinbaum2002logistic}
\bibfield{author}{\bibinfo{person}{David~G Kleinbaum}, \bibinfo{person}{K Dietz}, \bibinfo{person}{M Gail}, \bibinfo{person}{Mitchel Klein}, {and} \bibinfo{person}{Mitchell Klein}.} \bibinfo{year}{2002}\natexlab{}.
\newblock \bibinfo{booktitle}{\emph{Logistic regression}}.
\newblock \bibinfo{publisher}{Springer}.
\newblock


\bibitem[Lam et~al\mbox{.}(2006)]%
        {lam2006you}
\bibfield{author}{\bibinfo{person}{Shyong K~“Tony” Lam}, \bibinfo{person}{Dan Frankowski}, {and} \bibinfo{person}{John Riedl}.} \bibinfo{year}{2006}\natexlab{}.
\newblock \showarticletitle{Do you trust your recommendations? An exploration of security and privacy issues in recommender systems}. In \bibinfo{booktitle}{\emph{Emerging Trends in Information and Communication Security: International Conference, ETRICS 2006, Freiburg, Germany, June 6-9, 2006. Proceedings}}. Springer, \bibinfo{pages}{14--29}.
\newblock


\bibitem[Li et~al\mbox{.}(2016)]%
        {li2016data}
\bibfield{author}{\bibinfo{person}{Bo Li}, \bibinfo{person}{Yining Wang}, \bibinfo{person}{Aarti Singh}, {and} \bibinfo{person}{Yevgeniy Vorobeychik}.} \bibinfo{year}{2016}\natexlab{}.
\newblock \showarticletitle{Data poisoning attacks on factorization-based collaborative filtering}.
\newblock \bibinfo{journal}{\emph{Advances in neural information processing systems}}  \bibinfo{volume}{29} (\bibinfo{year}{2016}).
\newblock


\bibitem[Liu et~al\mbox{.}(2018)]%
        {liu2018stamp}
\bibfield{author}{\bibinfo{person}{Qiao Liu}, \bibinfo{person}{Yifu Zeng}, \bibinfo{person}{Refuoe Mokhosi}, {and} \bibinfo{person}{Haibin Zhang}.} \bibinfo{year}{2018}\natexlab{}.
\newblock \showarticletitle{STAMP: short-term attention/memory priority model for session-based recommendation}. In \bibinfo{booktitle}{\emph{Proceedings of the 24th ACM SIGKDD international conference on knowledge discovery \& data mining}}. \bibinfo{pages}{1831--1839}.
\newblock


\bibitem[Madry et~al\mbox{.}(2017)]%
        {madry2017towards}
\bibfield{author}{\bibinfo{person}{Aleksander Madry}, \bibinfo{person}{Aleksandar Makelov}, \bibinfo{person}{Ludwig Schmidt}, \bibinfo{person}{Dimitris Tsipras}, {and} \bibinfo{person}{Adrian Vladu}.} \bibinfo{year}{2017}\natexlab{}.
\newblock \showarticletitle{Towards deep learning models resistant to adversarial attacks}.
\newblock \bibinfo{journal}{\emph{arXiv preprint arXiv:1706.06083}} (\bibinfo{year}{2017}).
\newblock


\bibitem[McSherry and Mironov(2009)]%
        {mcsherry2009differentially}
\bibfield{author}{\bibinfo{person}{Frank McSherry} {and} \bibinfo{person}{Ilya Mironov}.} \bibinfo{year}{2009}\natexlab{}.
\newblock \showarticletitle{Differentially private recommender systems: Building privacy into the netflix prize contenders}. In \bibinfo{booktitle}{\emph{Proceedings of the 15th ACM SIGKDD international conference on Knowledge discovery and data mining}}. \bibinfo{pages}{627--636}.
\newblock


\bibitem[Miura et~al\mbox{.}(2021)]%
        {miura2021megex}
\bibfield{author}{\bibinfo{person}{Takayuki Miura}, \bibinfo{person}{Satoshi Hasegawa}, {and} \bibinfo{person}{Toshiki Shibahara}.} \bibinfo{year}{2021}\natexlab{}.
\newblock \showarticletitle{MEGEX: Data-free model extraction attack against gradient-based explainable AI}.
\newblock \bibinfo{journal}{\emph{arXiv preprint arXiv:2107.08909}} (\bibinfo{year}{2021}).
\newblock


\bibitem[Mnih and Salakhutdinov(2007)]%
        {mnih2007probabilistic}
\bibfield{author}{\bibinfo{person}{Andriy Mnih} {and} \bibinfo{person}{Russ~R Salakhutdinov}.} \bibinfo{year}{2007}\natexlab{}.
\newblock \showarticletitle{Probabilistic matrix factorization}.
\newblock \bibinfo{journal}{\emph{Advances in neural information processing systems}}  \bibinfo{volume}{20} (\bibinfo{year}{2007}).
\newblock


\bibitem[Nguyen~Thanh et~al\mbox{.}(2023)]%
        {nguyen2023poisoning}
\bibfield{author}{\bibinfo{person}{Toan Nguyen~Thanh}, \bibinfo{person}{Nguyen Duc~Khang Quach}, \bibinfo{person}{Thanh~Tam Nguyen}, \bibinfo{person}{Thanh~Trung Huynh}, \bibinfo{person}{Viet~Hung Vu}, \bibinfo{person}{Phi~Le Nguyen}, \bibinfo{person}{Jun Jo}, {and} \bibinfo{person}{Quoc Viet~Hung Nguyen}.} \bibinfo{year}{2023}\natexlab{}.
\newblock \showarticletitle{Poisoning GNN-based recommender systems with generative surrogate-based attacks}.
\newblock \bibinfo{journal}{\emph{ACM Transactions on Information Systems}} \bibinfo{volume}{41}, \bibinfo{number}{3} (\bibinfo{year}{2023}), \bibinfo{pages}{1--24}.
\newblock


\bibitem[Oh et~al\mbox{.}(2019)]%
        {oh2019towards}
\bibfield{author}{\bibinfo{person}{Seong~Joon Oh}, \bibinfo{person}{Bernt Schiele}, {and} \bibinfo{person}{Mario Fritz}.} \bibinfo{year}{2019}\natexlab{}.
\newblock \showarticletitle{Towards reverse-engineering black-box neural networks}.
\newblock \bibinfo{journal}{\emph{Explainable AI: Interpreting, Explaining and Visualizing Deep Learning}} (\bibinfo{year}{2019}), \bibinfo{pages}{121--144}.
\newblock


\bibitem[Orekondy et~al\mbox{.}(2019a)]%
        {orekondy2019knockoff}
\bibfield{author}{\bibinfo{person}{Tribhuvanesh Orekondy}, \bibinfo{person}{Bernt Schiele}, {and} \bibinfo{person}{Mario Fritz}.} \bibinfo{year}{2019}\natexlab{a}.
\newblock \showarticletitle{Knockoff nets: Stealing functionality of black-box models}. In \bibinfo{booktitle}{\emph{Proceedings of the IEEE/CVF conference on computer vision and pattern recognition}}. \bibinfo{pages}{4954--4963}.
\newblock


\bibitem[Orekondy et~al\mbox{.}(2019b)]%
        {orekondy2019prediction}
\bibfield{author}{\bibinfo{person}{Tribhuvanesh Orekondy}, \bibinfo{person}{Bernt Schiele}, {and} \bibinfo{person}{Mario Fritz}.} \bibinfo{year}{2019}\natexlab{b}.
\newblock \showarticletitle{Prediction poisoning: Towards defenses against dnn model stealing attacks}.
\newblock \bibinfo{journal}{\emph{arXiv preprint arXiv:1906.10908}} (\bibinfo{year}{2019}).
\newblock


\bibitem[Pazzani and Billsus(2007)]%
        {pazzani2007content}
\bibfield{author}{\bibinfo{person}{Michael~J Pazzani} {and} \bibinfo{person}{Daniel Billsus}.} \bibinfo{year}{2007}\natexlab{}.
\newblock \showarticletitle{Content-based recommendation systems}.
\newblock \bibinfo{journal}{\emph{The adaptive web: methods and strategies of web personalization}} (\bibinfo{year}{2007}), \bibinfo{pages}{325--341}.
\newblock


\bibitem[Rahman et~al\mbox{.}(2018)]%
        {rahman2018membership}
\bibfield{author}{\bibinfo{person}{Md~Atiqur Rahman}, \bibinfo{person}{Tanzila Rahman}, \bibinfo{person}{Robert Lagani{\`e}re}, \bibinfo{person}{Noman Mohammed}, {and} \bibinfo{person}{Yang Wang}.} \bibinfo{year}{2018}\natexlab{}.
\newblock \showarticletitle{Membership Inference Attack against Differentially Private Deep Learning Model.}
\newblock \bibinfo{journal}{\emph{Trans. Data Priv.}} \bibinfo{volume}{11}, \bibinfo{number}{1} (\bibinfo{year}{2018}), \bibinfo{pages}{61--79}.
\newblock


\bibitem[Rendle et~al\mbox{.}(2012)]%
        {rendle2012bpr}
\bibfield{author}{\bibinfo{person}{Steffen Rendle}, \bibinfo{person}{Christoph Freudenthaler}, \bibinfo{person}{Zeno Gantner}, {and} \bibinfo{person}{Lars Schmidt-Thieme}.} \bibinfo{year}{2012}\natexlab{}.
\newblock \showarticletitle{BPR: Bayesian personalized ranking from implicit feedback}.
\newblock \bibinfo{journal}{\emph{arXiv preprint arXiv:1205.2618}} (\bibinfo{year}{2012}).
\newblock


\bibitem[Schafer et~al\mbox{.}(2001)]%
        {schafer2001commerce}
\bibfield{author}{\bibinfo{person}{J~Ben Schafer}, \bibinfo{person}{Joseph~A Konstan}, {and} \bibinfo{person}{John Riedl}.} \bibinfo{year}{2001}\natexlab{}.
\newblock \showarticletitle{E-commerce recommendation applications}.
\newblock \bibinfo{journal}{\emph{Data mining and knowledge discovery}}  \bibinfo{volume}{5} (\bibinfo{year}{2001}), \bibinfo{pages}{115--153}.
\newblock


\bibitem[Shani and Gunawardana(2011)]%
        {shani2011evaluating}
\bibfield{author}{\bibinfo{person}{Guy Shani} {and} \bibinfo{person}{Asela Gunawardana}.} \bibinfo{year}{2011}\natexlab{}.
\newblock \showarticletitle{Evaluating recommendation systems}.
\newblock \bibinfo{journal}{\emph{Recommender systems handbook}} (\bibinfo{year}{2011}), \bibinfo{pages}{257--297}.
\newblock


\bibitem[Shokri et~al\mbox{.}(2017)]%
        {shokri2017membership}
\bibfield{author}{\bibinfo{person}{Reza Shokri}, \bibinfo{person}{Marco Stronati}, \bibinfo{person}{Congzheng Song}, {and} \bibinfo{person}{Vitaly Shmatikov}.} \bibinfo{year}{2017}\natexlab{}.
\newblock \showarticletitle{Membership inference attacks against machine learning models}. In \bibinfo{booktitle}{\emph{2017 IEEE symposium on security and privacy (SP)}}. IEEE, \bibinfo{pages}{3--18}.
\newblock


\bibitem[Takemura et~al\mbox{.}(2020)]%
        {takemura2020model}
\bibfield{author}{\bibinfo{person}{Tatsuya Takemura}, \bibinfo{person}{Naoto Yanai}, {and} \bibinfo{person}{Toru Fujiwara}.} \bibinfo{year}{2020}\natexlab{}.
\newblock \showarticletitle{Model extraction attacks on recurrent neural networks}.
\newblock \bibinfo{journal}{\emph{Journal of Information Processing}}  \bibinfo{volume}{28} (\bibinfo{year}{2020}), \bibinfo{pages}{1010--1024}.
\newblock


\bibitem[Tram{\`e}r et~al\mbox{.}(2016)]%
        {tramer2016stealing}
\bibfield{author}{\bibinfo{person}{Florian Tram{\`e}r}, \bibinfo{person}{Fan Zhang}, \bibinfo{person}{Ari Juels}, \bibinfo{person}{Michael~K Reiter}, {and} \bibinfo{person}{Thomas Ristenpart}.} \bibinfo{year}{2016}\natexlab{}.
\newblock \showarticletitle{Stealing Machine Learning Models via Prediction APIs.}. In \bibinfo{booktitle}{\emph{USENIX security symposium}}, Vol.~\bibinfo{volume}{16}. \bibinfo{pages}{601--618}.
\newblock


\bibitem[Wang and Gong(2018)]%
        {wang2018stealing}
\bibfield{author}{\bibinfo{person}{Binghui Wang} {and} \bibinfo{person}{Neil~Zhenqiang Gong}.} \bibinfo{year}{2018}\natexlab{}.
\newblock \showarticletitle{Stealing hyperparameters in machine learning}. In \bibinfo{booktitle}{\emph{2018 IEEE symposium on security and privacy (SP)}}. IEEE, \bibinfo{pages}{36--52}.
\newblock


\bibitem[Wang et~al\mbox{.}(2022b)]%
        {wang2022black}
\bibfield{author}{\bibinfo{person}{Yixu Wang}, \bibinfo{person}{Jie Li}, \bibinfo{person}{Hong Liu}, \bibinfo{person}{Yan Wang}, \bibinfo{person}{Yongjian Wu}, \bibinfo{person}{Feiyue Huang}, {and} \bibinfo{person}{Rongrong Ji}.} \bibinfo{year}{2022}\natexlab{b}.
\newblock \showarticletitle{Black-box dissector: Towards erasing-based hard-label model stealing attack}. In \bibinfo{booktitle}{\emph{Computer Vision--ECCV 2022: 17th European Conference, Tel Aviv, Israel, October 23--27, 2022, Proceedings, Part V}}. Springer, \bibinfo{pages}{192--208}.
\newblock


\bibitem[Wang et~al\mbox{.}(2022a)]%
        {wang2022debiasing}
\bibfield{author}{\bibinfo{person}{Zihan Wang}, \bibinfo{person}{Na Huang}, \bibinfo{person}{Fei Sun}, \bibinfo{person}{Pengjie Ren}, \bibinfo{person}{Zhumin Chen}, \bibinfo{person}{Hengliang Luo}, \bibinfo{person}{Maarten de Rijke}, {and} \bibinfo{person}{Zhaochun Ren}.} \bibinfo{year}{2022}\natexlab{a}.
\newblock \showarticletitle{Debiasing Learning for Membership Inference Attacks Against Recommender Systems}. In \bibinfo{booktitle}{\emph{Proceedings of the 28th ACM SIGKDD Conference on Knowledge Discovery and Data Mining}}. \bibinfo{pages}{1959--1968}.
\newblock


\bibitem[Wei et~al\mbox{.}(2007)]%
        {wei2007survey}
\bibfield{author}{\bibinfo{person}{Kangning Wei}, \bibinfo{person}{Jinghua Huang}, {and} \bibinfo{person}{Shaohong Fu}.} \bibinfo{year}{2007}\natexlab{}.
\newblock \showarticletitle{A survey of e-commerce recommender systems}. In \bibinfo{booktitle}{\emph{2007 international conference on service systems and service management}}. IEEE, \bibinfo{pages}{1--5}.
\newblock


\bibitem[Wu et~al\mbox{.}(2022)]%
        {wu2022model}
\bibfield{author}{\bibinfo{person}{Bang Wu}, \bibinfo{person}{Xiangwen Yang}, \bibinfo{person}{Shirui Pan}, {and} \bibinfo{person}{Xingliang Yuan}.} \bibinfo{year}{2022}\natexlab{}.
\newblock \showarticletitle{Model extraction attacks on graph neural networks: Taxonomy and realisation}. In \bibinfo{booktitle}{\emph{Proceedings of the 2022 ACM on Asia Conference on Computer and Communications Security}}. \bibinfo{pages}{337--350}.
\newblock


\bibitem[Wu et~al\mbox{.}(2021)]%
        {wu2021triple}
\bibfield{author}{\bibinfo{person}{Chenwang Wu}, \bibinfo{person}{Defu Lian}, \bibinfo{person}{Yong Ge}, \bibinfo{person}{Zhihao Zhu}, {and} \bibinfo{person}{Enhong Chen}.} \bibinfo{year}{2021}\natexlab{}.
\newblock \showarticletitle{Triple adversarial learning for influence based poisoning attack in recommender systems}. In \bibinfo{booktitle}{\emph{Proceedings of the 27th ACM SIGKDD Conference on Knowledge Discovery \& Data Mining}}. \bibinfo{pages}{1830--1840}.
\newblock


\bibitem[Yu et~al\mbox{.}(2020)]%
        {yu2020cloudleak}
\bibfield{author}{\bibinfo{person}{Honggang Yu}, \bibinfo{person}{Kaichen Yang}, \bibinfo{person}{Teng Zhang}, \bibinfo{person}{Yun-Yun Tsai}, \bibinfo{person}{Tsung-Yi Ho}, {and} \bibinfo{person}{Yier Jin}.} \bibinfo{year}{2020}\natexlab{}.
\newblock \showarticletitle{CloudLeak: Large-Scale Deep Learning Models Stealing Through Adversarial Examples.}. In \bibinfo{booktitle}{\emph{NDSS}}.
\newblock


\bibitem[Yuan et~al\mbox{.}(2023)]%
        {yuan2023interaction}
\bibfield{author}{\bibinfo{person}{Wei Yuan}, \bibinfo{person}{Chaoqun Yang}, \bibinfo{person}{Quoc Viet~Hung Nguyen}, \bibinfo{person}{Lizhen Cui}, \bibinfo{person}{Tieke He}, {and} \bibinfo{person}{Hongzhi Yin}.} \bibinfo{year}{2023}\natexlab{}.
\newblock \showarticletitle{Interaction-level Membership Inference Attack Against Federated Recommender Systems}.
\newblock \bibinfo{journal}{\emph{arXiv preprint arXiv:2301.10964}} (\bibinfo{year}{2023}).
\newblock


\bibitem[Yue et~al\mbox{.}(2021)]%
        {yue2021black}
\bibfield{author}{\bibinfo{person}{Zhenrui Yue}, \bibinfo{person}{Zhankui He}, \bibinfo{person}{Huimin Zeng}, {and} \bibinfo{person}{Julian McAuley}.} \bibinfo{year}{2021}\natexlab{}.
\newblock \showarticletitle{Black-box attacks on sequential recommenders via data-free model extraction}. In \bibinfo{booktitle}{\emph{Proceedings of the 15th ACM Conference on Recommender Systems}}. \bibinfo{pages}{44--54}.
\newblock


\bibitem[Zhan et~al\mbox{.}(2010)]%
        {zhan2010privacy}
\bibfield{author}{\bibinfo{person}{Justin Zhan}, \bibinfo{person}{Chia-Lung Hsieh}, \bibinfo{person}{I-Cheng Wang}, \bibinfo{person}{Tsan-Sheng Hsu}, \bibinfo{person}{Churn-Jung Liau}, {and} \bibinfo{person}{Da-Wei Wang}.} \bibinfo{year}{2010}\natexlab{}.
\newblock \showarticletitle{Privacy-preserving collaborative recommender systems}.
\newblock \bibinfo{journal}{\emph{IEEE Transactions on Systems, Man, and Cybernetics, Part C (Applications and Reviews)}} \bibinfo{volume}{40}, \bibinfo{number}{4} (\bibinfo{year}{2010}), \bibinfo{pages}{472--476}.
\newblock


\bibitem[Zhang et~al\mbox{.}(2021a)]%
        {zhang2021membership}
\bibfield{author}{\bibinfo{person}{Minxing Zhang}, \bibinfo{person}{Zhaochun Ren}, \bibinfo{person}{Zihan Wang}, \bibinfo{person}{Pengjie Ren}, \bibinfo{person}{Zhunmin Chen}, \bibinfo{person}{Pengfei Hu}, {and} \bibinfo{person}{Yang Zhang}.} \bibinfo{year}{2021}\natexlab{a}.
\newblock \showarticletitle{Membership inference attacks against recommender systems}. In \bibinfo{booktitle}{\emph{Proceedings of the 2021 ACM SIGSAC Conference on Computer and Communications Security}}. \bibinfo{pages}{864--879}.
\newblock


\bibitem[Zhang and Yin(2022)]%
        {zhang2022comprehensive}
\bibfield{author}{\bibinfo{person}{Shijie Zhang} {and} \bibinfo{person}{Hongzhi Yin}.} \bibinfo{year}{2022}\natexlab{}.
\newblock \showarticletitle{Comprehensive Privacy Analysis on Federated Recommender System against Attribute Inference Attacks}.
\newblock \bibinfo{journal}{\emph{arXiv preprint arXiv:2205.11857}} (\bibinfo{year}{2022}).
\newblock


\bibitem[Zhang et~al\mbox{.}(2021b)]%
        {zhang2021graph}
\bibfield{author}{\bibinfo{person}{Shijie Zhang}, \bibinfo{person}{Hongzhi Yin}, \bibinfo{person}{Tong Chen}, \bibinfo{person}{Zi Huang}, \bibinfo{person}{Lizhen Cui}, {and} \bibinfo{person}{Xiangliang Zhang}.} \bibinfo{year}{2021}\natexlab{b}.
\newblock \showarticletitle{Graph embedding for recommendation against attribute inference attacks}. In \bibinfo{booktitle}{\emph{Proceedings of the Web Conference 2021}}. \bibinfo{pages}{3002--3014}.
\newblock


\bibitem[Zhang et~al\mbox{.}(2021c)]%
        {zhang2021adversarial}
\bibfield{author}{\bibinfo{person}{Xingwei Zhang}, \bibinfo{person}{Xiaolong Zheng}, {and} \bibinfo{person}{Wenji Mao}.} \bibinfo{year}{2021}\natexlab{c}.
\newblock \showarticletitle{Adversarial perturbation defense on deep neural networks}.
\newblock \bibinfo{journal}{\emph{ACM Computing Surveys (CSUR)}} \bibinfo{volume}{54}, \bibinfo{number}{8} (\bibinfo{year}{2021}), \bibinfo{pages}{1--36}.
\newblock


\bibitem[Zhu et~al\mbox{.}(2023)]%
        {zhu2023membership}
\bibfield{author}{\bibinfo{person}{Zhihao Zhu}, \bibinfo{person}{Chenwang Wu}, \bibinfo{person}{Rui Fan}, \bibinfo{person}{Defu Lian}, {and} \bibinfo{person}{Enhong Chen}.} \bibinfo{year}{2023}\natexlab{}.
\newblock \showarticletitle{Membership Inference Attacks Against Sequential Recommender Systems}. In \bibinfo{booktitle}{\emph{Proceedings of the ACM Web Conference 2023}}. \bibinfo{pages}{1208--1219}.
\newblock


\end{thebibliography}

\clearpage
\appendix

\section{APPENDIX}
\label{appendix}

\begin{table}[]
\centering
\caption{Statistics of datasets}
\begin{tabular}{l|rrrr}
\hline
Dataset      & \#Users & \#Items & \#Interactions &Sparsity \\
\hline
ML-1M        & 6040    & 3706    & 1000209        & 95.53\% \\ 
Steam        & 25387   & 4081    & 313856         & 99.70\% \\ 
Ta-feng      & 19451   & 10480   & 630767         & 99.69\% \\
\hline
\end{tabular}
\label{datasets}
\end{table}

\subsection{Setup}
\label{Setup}
\subsubsection{Dataset} 
We evaluate the attack methods on three real-world datasets, including ML-1M(MovieLens-1M), Ta-feng, and Steam.
We remove users and items with less than 5 interaction records.
The statistics of the filtered datasets are summarized in Table~\ref{datasets}.
We divide each dataset evenly into two disjoint subsets: a target dataset and an auxiliary dataset. 
We assume that the attacker has access to 10\% of the target dataset for training the clone model.
\begin{itemize}[leftmargin = 10 pt]
\item \textbf{ML-1M}~\cite{harper2015movielens} is a widely used dataset in recommender systems, comprising over one million movie ratings by users on a scale of 1 to 5.
Additionally, ML-1M provides basic information about its users and movies, such as age, gender, and movie genres.
\item \textbf{Ta-feng}\footnote{https://www.kaggle.com/datasets/chiranjivdas09/ta-feng-grocery-dataset} contains transactional data from a Chinese grocery store spanning from November 2000 to February 2001. 
It keeps a record of users' purchase history as well as the purchase time.
\item \textbf{Steam}\footnote{https://www.kaggle.com/datasets/tamber/steam-video-games} collects player reviews and game information on the Steam platform, including the time that players purchased games and the total amount of time they spent playing those games.
\end{itemize}

\subsubsection{Recommender Systems}
To evaluate the effectiveness of various attack algorithms, we conduct experiments against three classic recommender systems, namely BPR(Bayesian Personalized Ranking), NCF(Neural Collaborative Filtering), and LMF(Logistic Matrix Factorization).
\begin{itemize}[leftmargin = 10 pt]
\item \textbf{BPR}~\cite{rendle2012bpr} is a commonly used recommender system that utilizes users’ interaction history. It assumes that the preference behavior between each user is independent of each other and models users by analyzing their interaction records. 
\item \textbf{NCF}\cite{he2017neural} proposes a recommender system based on neural networks. 
It makes use of neural networks to predict user ratings for items, as opposed to the traditional recommender systems, which derive user ratings for items via inner product operations. 
\item \textbf{LMF}\cite{johnson2014logistic} is a probabilistic recommender system in which the value of the latent factor matrices of users and items are assumed to follow a Gaussian distribution. 
Utilizing a probabilistic approach, during the training process, LMF maximizes the likelihood of users being recommended for items they have interacted with. 
\end{itemize}
In order to assure a fair comparison with prior research, we also analyzed the attack performance against three sequential recommender systems, including SASRec, STAMP, and GRU4Rec. 
\begin{itemize}[leftmargin = 10 pt]
    \item \textbf{SASRec}~\cite{kang2018self} proposes a sequential recommendation algorithm based on self-attention mechanism.
    \item \textbf{STAMP}~\cite{liu2018stamp} captures both users' general interests and their current preferences through a short-term memory priority.
    \item \textbf{GRU4Rec}~\cite{hidasi2015session} utilizes Gated Recurrent Units (GRU) to model the click sequences of each user.
\end{itemize}

\subsubsection{Attack Algorithms}
In this part, we describe the baselines used in our experiment, including PTD(\textbf{P}artial \textbf{T}arget \textbf{D}ata) and QSD(\textbf{Q}uery with \textbf{S}ynthetic \textbf{D}ata).
\begin{itemize}[leftmargin = 10 pt]
\item \textbf{PTD}~\cite{li2016data, nguyen2023poisoning} : Most adversarial attacks against recommender systems assume that they have access to plenty of the target data. They train a clone model with the available target data as we did in Section~\ref{PTD}. The clone model is further utilized to implement adversarial attacks against the target model. The following results are generated based on the assumption that the attackers possess access to target data of equivalent size to that which we have at our disposal. 
\item \textbf{QSD}~\cite{yue2021black} : The attacker trains the clone model using synthetic data rather than real data in QSD.
The attacker exploits the auto-regressive nature of sequential recommender systems to generate fixed-length user interactions gradually. 
The attacker makes use of the permission to query the target model in order to guide the process of data production, which ultimately results in the acquisition of more realistic synthetic(fake) data. In the next step of the attack, the attacker will collect information from the recommendations supplied by the target model for synthetic data in order to train a more accurate clone model.
However, the application of this data generation method is limited by the unique autoregressive characteristics of sequential recommendation models and the large number of queries required. 
To evaluate its performance on other recommender systems, we remove the synthetic data generation and assume that it can obtain partial target data. 
Then we extract the information from the target model's recommendations for these target data by query permission as ~\cite{yue2021black} did.
It should be emphasized that, while QSD has some of the target data, we refer to its provenance~\cite{yue2021black} and only use the recommendation list from the query access to train the clone model, rather than the interaction of the target data.
\end{itemize}

\subsubsection{Implementation Details}
We employ Adam optimizer with a learning rate of 0.001 to train recommendation models.
The batch size is set to 2048.
The architecture of recommendation models is basically the same as that in the original papers. 
Also, we hypothesize that the attacker can access 10\% of the target dataset. 
Unlike QSD~\cite{yue2021black}, which fixes the size of the candidate items to 100, we take all items (items size is 3706 in ML-1M) as candidate items.
We assume that the target model produces a recommendation list with a length of 50.
We set the value of margin in margin hinge loss to 0.5, and the length of user and item embeddings to 64.

\begin{table*}[]
\centering
\caption{Attack performance with various stealing loss functions}
\begin{tabular}{l|ccc|ccc|ccc}
\noalign{\smallskip}\hline\noalign{\smallskip}
\multirow{2}{*}{Loss Function} & \multicolumn{3}{c|}{BPR}  & \multicolumn{3}{c|}{NCF}  & \multicolumn{3}{c}{LMF} \\   \noalign{\smallskip}\cline{2-10} \noalign{\smallskip}
                                 & QSD    & PTQ    & PTAQ   & QSD    & PTQ    & PTAQ   & QSD      & PTQ      & PTAQ     \\ \noalign{\smallskip}\hline\noalign{\smallskip}
Hinge+Hinge                      & 0.0533 & 0.5040 & 0.6387 & \textbf{0.2727} & \textbf{0.3633} & 0.4427 & \textbf{0.0487}   & 0.3000   & 0.5093   \\
Hinge+BPR                        & \textbf{0.0567} & 0.5027 & 0.6393 & 0.2113 & \textbf{0.3633} & \textbf{0.4433} & 0.0440   & 0.3040    & 0.5033   \\
\textbf{BPR+Hinge}               & 0.0540 & \textbf{0.5107} & \textbf{0.6506} & 0.2413 & \textbf{0.3633} & 0.4407 & 0.0480   & 0.2873   & \textbf{0.5627}    \\
BPR+BPR                          & 0.0547 & \textbf{0.5107} & 0.6447 & 0.2540 & 0.3479 & 0.4360 & \textbf{0.0487}   & \textbf{0.3100}   & 0.5100     \\
\noalign{\smallskip}\hline
\end{tabular}
\label{stealloss}
\end{table*}

\begin{figure*}[h]
  \centering
  \includegraphics[width=0.8\linewidth]{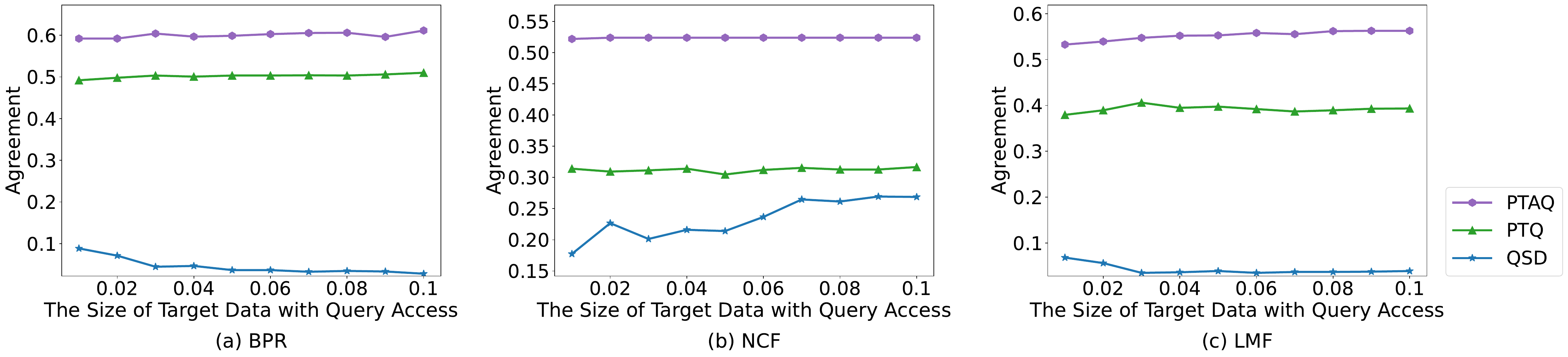}
  \caption{Attack performance with different query budgets on ML-1M.}
  \label{query_rate}
\end{figure*}

\begin{table}[]
\centering
\caption{Attack performance on sequential recommender systems on ML-1M.}
\begin{tabular}{l|ccc}
\noalign{\smallskip}\hline\noalign{\smallskip}
Algorithm                  & SASRec & STAMP  & GRU4Rec \\
\noalign{\smallskip}\hline\noalign{\smallskip}
QSD5                       & 0.0528 & 0.0718 & 0.0616  \\
QSD10                      & 0.0551 & 0.2961 & 0.1007  \\
QSD15                      & 0.0616 & 0.0970 & 0.0944  \\
QSD20                      & 0.0623 & 0.3997 & 0.0964  \\
QSD25                      & 0.4902 & 0.4318 & 0.0974  \\
QSD30                      & \textbf{0.5128} & \textbf{0.4436} & \textbf{0.1390}  \\
\noalign{\smallskip}\hline\noalign{\smallskip}
PTD                        & 0.5679 & 0.3948 & 0.4433  \\
PTQ                        & 0.5856 & 0.3948 & 0.4433  \\
PTAQ                       & \textbf{0.6173} & \textbf{0.4705} & \textbf{0.5013}  \\
\noalign{\smallskip}\hline
\end{tabular}
\label{seq results}
\end{table}

\subsection{Influence of Stealing Loss Function}
In this section, we explore various stealing loss functions and evaluated their effect on the efficacy of the attack algorithm. 
The performance evaluation of QSD, PTQ, and PTAQ is conducted under six distinct stealing functions, which are exclusively employed in attack methods that utilize query feedback. The results of this evaluation have been reported in Table~\ref{stealloss}. 
BPR+Hinge indicates that we used the BPR loss to extract the ranking information of the recommendation list and the margin hinge loss to extract the recommended item information of the recommendation list. 
In the previous experiments, PTQ and PTAQ used \textbf{BPR+Hinge} as the stealing loss function, and QSD used \textbf{Hinge+Hinge} as the stealing loss function.
Our findings indicate that the efficacy of the attack approach remains largely unaffected by the stealing function. A reasonably effective stealing function is adequate for leveraging the knowledge obtained from the recommendations. 

\subsection{Impact of Query Budgets}
\label{Attack Performance with Few Query Budgets}
Figure~\ref{query_rate} analyzed attack performances with respect to the varying size of target data with query access. 
From the experimental results, we find that PTAQ and PTQ are less affected by the number of queries because they make appropriate use of the user-item interaction of the target(auxiliary) data. However, QSD is greatly affected by the number of queries. As mentioned earlier, QSD generates a mass of synthetic data and makes corresponding queries. Its performance is intricately linked to the number of inquiries made toward the target model. To make a more reliable comparison with QSD, we analyzed the performance of different attack algorithms on sequential recommendation models on the ML-1M dataset. We denote QSD with \textbf{k times} the number of queries as our methods as QSDk. From the observations in Table~\ref{seq results}, we draw the following conclusions: 
\begin{itemize}[leftmargin = 10 pt]
    \item The efficacy of QSD is significantly influenced by the number of inquiries made to the target model. In scenarios where the query count is limited, the QSD may not present a significant challenge to the target model. However, when QSD is granted the ability to execute numerous queries, its efficiency can undergo a significant improvement. The experimental results indicate that QSD30 exhibits attack performance of 51.28\% and 44.36\% against SASRec and STAMP, respectively. 
    In the absence of target data, the attack efficacy of QSD may match or exceed that of PTD. However, making a large number of queries is not a justifiable approach. The utilization of auxiliary data can potentially mitigate the query volume needed for sequential recommendation models. The proposed methodology, PTAQ, demonstrates superior attack efficacy in comparison to QSD, while utilizing significantly fewer queries. The rationale for this experimental finding is that we \textbf{make efficient use of the user-item interactions} from the target data and auxiliary data.
\end{itemize}

\end{document}